\documentclass[10pt]{article}
\usepackage{latexsym}
\usepackage{amssymb}
\usepackage{amsmath}
\usepackage{amscd}
\usepackage{amsthm}
\usepackage[left=2.5cm,top=1cm,right=2.5cm,bottom=2cm]{geometry}

\usepackage[dvips]{graphicx}
\usepackage{hyperref}
\begin{document}
\begin{center}
\large{\bf {LRS Bianchi Type-II Massive String Cosmological Models in General Relativity}} \\
\vspace{10mm}
\normalsize{Anil Kumar Yadav $^1$, Anirudh Pradhan $^2$ and Archana Singh $^3$}\\
\vspace{5mm} \normalsize{$^{1}$Department of Physics, Anand Engineering College, 
Agra-282 007, India \\
E-mail : abanilyadav@yahoo.co.in, akyadav@imsc.res.in} \\
\vspace{5mm} \normalsize{$^{2}$Department of Mathematics, Hindu Post-graduate College, Zamania-232 331, 
Ghazipur, India \\
E-mail : acpradhan@yahoo.com, apradhan@imsc.res.in} \\
\vspace{5mm}
\normalsize{$^{3}$Department of Mathematics, Post-graduate College, Ghazipur-221 001, India}\\
\end{center}
\vspace{5mm}
\begin{abstract}
The present study deals with locally rotationally symmetric (LRS) Bianchi type II cosmological models representing 
massive string. The energy-momentum tensor for such string as formulated by Letelier (1983) is used to construct 
massive string cosmological models for which we assume that the expansion ($\theta$) in the models is proportional 
to the shear ($\sigma$). This condition leads to $A = B^{m}$, where A and B are the metric coefficients and m is 
constant. We have derived two types of models depending on different values of m i.e. for $m\neq \sqrt{2}$ and 
$m = \sqrt{2}$ respectively. For suitable choice of constant m (i.e. for $m = \frac{1}{2}$), it is observed that 
in early stage of the evolution of the universe, the universe is dominated by strings in both cases. Our models are 
in accelerating phase which is consistent to the recent observations of type Is supernovae. Some physical and geometric 
behaviour of the models are also discussed.
\end{abstract}
\section{Introduction}
The large scale matter distribution in the observable universe, largely manifested in the form of discrete structures, 
does not exhibit homogeneity of a high order. In contrast, the cosmic background radiation (CMB), which is significant 
in the microwave region, is extremely homogeneous, however, recent space investigations detect anisotropy in the CMB. 
The observations from cosmic background explorer’s differential microwave radiometer (COBE-DMR) have detected and 
measured CMB anisotropies in different angular scales. These anisotropies are supposed to hide in their fold the entire 
history of cosmic evolution dating back to the recombination epoch and are being considered as indicative of the 
geometry and the content of the universe. More about CMB anisotropy is expected to be uncovered by the investigations 
of Wilkinson Microwave Anisotropy Probe (WMAP). There is widespread consensus among cosmologists that CMB anisotropies 
in small angular scales have the key to the formation of discrete structures. Our interest is in observed CMB 
anisotropies in the large angular scales and we intend to attribute it to the anisotropy of the spatial cosmic geometry. 
Bianchi type space-times exhibit spatial homogeneity and anisotropy. In this paper we assume that the geometry of the 
universe is that of a locally rotationally symmetric (LRS) Bianchi type II space-time.\\

In recent years, there has been considerable interest in string cosmology. Cosmic strings are topologically stable 
objects which might be found during a phase transition in the early universe (Kibble \cite{ref1}). Cosmic strings 
play an important role in the study of the early universe. These arise during the phase transition after the big 
bang explosion as the temperature goes down below some critical temperature as predicted by grand unified theories 
(Zel'dovich et al. \cite{ref2,ref3}; Kibble \cite{ref1,ref4}; Everett \cite{ref5}; Vilenkin \cite{ref6}). It is believed 
that cosmic strings give rise to density perturbations which lead to the formation of galaxies (Zel'dovich 
\cite {ref7}). These cosmic strings have stress-energy and couple to the gravitational field. Therefore it is 
interesting to study the gravitational effects that arise from strings. The pioneering work in the formulation of the
energy-momentum tensor for classical massive strings was done by Letelier \cite{ref8} who considered the massive 
strings to be formed by geometric strings with particle attached along its extension. Letelier \cite{ref9} first used 
this idea in obtaining cosmological solutions in Bianchi I and Kantowski-Sachs space-times. Stachel \cite{ref10} has 
studied massive string.\\

Bali et al. \cite{ref11}$-$\cite{ref17} have obtained Bianchi types-I, III and IX string cosmological models in general 
relativity. Yadav et al. \cite{ref18} have studied some Bianchi type-I viscous fluid string cosmological models with 
magnetic field. Recently Wang \cite{ref19}$-$\cite{ref22} has also discussed LRS Bianchi type-I and Bianchi type-III 
cosmological models for a cloud string with bulk viscosity. Saha {\it et al.} \cite{ref23,ref24} have studied Bianchi 
type-I cosmological models in presence of magnetic flux. Yadav {\it et al.} \cite {ref25} have obtained 
cylindrically symmetric inhomogeneous universe with a cloud of strings. Baysal 
{\it et al.} \cite {ref26} have investigated the behaviour of a string in the cylindrically symmetric inhomogeneous 
universe. Yavuz {\it et al.} \cite {ref27} have examined charged strange quark matter attached to the string cloud in 
the spherically symmetric space-time admitting one-parameter group of conformal motion. Kaluza-Klein cosmological 
solutions are obtained by Yilmaz \cite {ref28}. Singh \& Singh \cite {ref29}, and Singh \cite {ref30,ref31} have 
studied string cosmological models in different symetry and Bianch space-times. Reddy \cite{ref32,ref33}, Reddy 
{\it et al.} \cite{ref34}$-$\cite{ref36}, Rao et al. \cite{ref37}$-$\cite{ref40}, Pradhan \cite{ref41,ref42}, and 
Pradhan {\it et al.} \cite{ref43}$-$ \cite{ref45} have studied string cosmological models in different contexts. 
Recently, Tripathi {\it et al.} \cite{ref46,ref47} have obtained cosmic strings with bulk viscosity.\\

The present day universe is satisfactorily described by homogeneous and isotropic models given by the FRW space-time. 
The universe in a smaller scale is neither homogeneous nor isotropic nor do we expect the Universe in its early stages 
to have these properties . Homogeneous and anisotropic cosmological models have been widely studied in the framework 
of General Relativity in the search of a realistic picture of the universe in its early stages. Although these are more 
restricted than the inhomogeneous models which explain a number of observed phenomena quite satisfactorily. Bianchi
type-II space-time has a fundamental role in constructing cosmological models suitable for describing the early stages 
of evolution of universe. Asseo and Sol \cite{ref48} emphasized the importance of Bianchi type-II universe. A spatially 
homogeneous Bianchi model necessarily has a three-dimensional group, which acts simply transitively on space-like 
three-dimensional orbits. Here we confine ourselves to a locally rotationally symmetric (LRS) model of Bianchi type-II.
This model is characterized by three metric functions $R_{1}(t)$, $R_{2}(t)$ and $R_{3}(t)$ such that $R_{1} = R_{2} 
\ne R_{3}$. The metric functions are functions of time only. (For non-LRS Bianchi metrics we have $R_{1} \ne R_{2} 
\ne R_{3}$). For LRS Bianchi type-II metric, Einstein's field equations reduce, in the case of perfect fluid 
distribution of matter, to three nonlinear differential equations. \\

Roy and Banerjee \cite{ref49} have dealt with LRS cosmological models of Bianchi type-II representing clouds of 
geometrical as well as massive strings. Wang \cite{ref50} studied the Letelier model in the context of LRS Bianchi 
type-II space-time. Belinchon \cite{ref51,ref52} studied Bianchi type-II space-time in connection with massive cosmic 
string and perfect fluid models with time varying constants under the self-similarity approach respectively. Recently, 
Pradhan, Amirhashchi and Yadav \cite{ref53} and Amirhashchi and Zainuddin \cite{ref54} obtained LRS Bianchi type-II 
cosmological models with perfect fluid distribution of matter and string dust respectively. Recently, Pradhan 
{\it et al.} \cite{ref55} obtained an LRS Bianchi type-II massive string cosmological model in general relativity. The 
strings that form the cloud are massive strings instead of geometric strings. Each massive string is formed by a 
geometric string with particles attached along its extension. Hence, the string that form the cloud are generalization 
of Takabayasi's relativistic model of strings (called p-string). This is simplest model wherein we have particles and 
strings together. Motivated by the situations discussed above, in this paper, we have revisited the solution of Pradhan 
{\it et al.} \cite{ref55} and find out two types of general solutions for LRS Bianchi type-II space-time for a cloud of 
strings which are different from the other author's solutions. Our solutions generalize the solutions recently obtained 
by Pradhan {\it et al.} \cite{ref55}. The paper is organized as follows. The metric and the field equations are presented 
in Section 2. In Section 3, we deal with two types of exact solutions of the field equations with cloud of strings. In 
Subsections 3.1 and 3.2, we deal with two cases, i.e., for $m \ne \sqrt{2}$ and $m = \sqrt{2}$ respectively. In Section 
4, we describe some physical and geometric properties of the both models. Finally, in Section 5, concluding remarks 
have been given. \\
\section{The Metric and Field  Equations}
We consider the LRS Bianchi type II metric in the form
\begin{equation}
\label{eq1}
ds^{2} =  - dt^{2} + B^{2}(dx + zdy)^{2} + A^{2} (dy^{2} + dz^{2}),
\end{equation}
where A and B are functions of t only. The energy momentum tensor for a cloud of string is taken as
\begin{equation}
\label{eq2}
T^{j}_{i} = \rho v_{i}v^{j} - \lambda x_{i}x^{j},
\end{equation}
where $v_{i}$ and $x_{i}$ satisfy condition
\begin{equation}
\label{eq3}
v^{i} v_{i} = - x^{i} x_{i} = -1, \, \, \, v^{i} x_{i} = 0,
\end{equation}
where $\rho$ is the proper energy density for a cloud string with particles attached to them, $\lambda$ 
is the string tension density, $v^{i}$ the four-velocity of the particles, and $x^{i}$ is a unit space-like 
vector representing the direction of string. In a co-moving co-ordinate system, we have
\begin{equation}
\label{eq4}
v^{i} = (0, 0, 0, 1), \, \, \, x^{i} = \left(\frac{1}{B}, 0, 0, 0 \right).
\end{equation}
If the particle density of the configuration is denoted by
$\rho_{p}$, then we have
\begin{equation}
\label{eq5}
\rho = \rho_{p} + \lambda.
\end{equation}
The Einstein's field equations (in geometrized units $ 8\pi G = c = 1 $)
\begin{equation}
\label{eq6}
R^{j}_{i} - \frac{1}{2} R g^{j}_{i} = - T^{j}_{i},
\end{equation}
for the line-element (\ref{eq1}) lead to
\begin{equation}
\label{eq7}
\frac{2\ddot{A}}{A} + \frac{\dot{A}^{2}}{A^{2}} - \frac{3}{4} \frac{B^{2}}{A^{4}} = - \lambda,
\end{equation}
\begin{equation}
\label{eq8}
\frac{\ddot{A}}{A} + \frac{\ddot{B}}{B} + \frac{\dot{A}\dot{B}}{AB} + \frac{1}{4} \frac{B^{2}}{A^{4}} = 0,
\end{equation}
\begin{equation}
\label{eq9}
\frac{2\dot{A}\dot{B}}{AB} + \frac{\dot{A}^{2}}{A^{2}} - \frac{1}{4} \frac{B^{2}}{A^{4}}  =  \rho,
\end{equation}
where an over dot stands for the first and double over dot for the second derivative with respect to $t$.
The particle density $\rho_{p} = (\rho -\lambda)$ is given by
\begin{equation}
\label{eq10}
\rho_{p} = \frac{2\dot{A}\dot{B}}{AB} + \frac{1}{2} \frac{B^{2}}{A^{4}} - \frac{2\ddot{A}}{A}.
\end{equation}
The average scale factor $a$ of LRS Bianchi type-II model is defined as
\begin{equation}
\label{eq11} a = (A^{2}B)^{\frac{1}{3}}.
\end{equation}
A volume scale factor V is given by
\begin{equation}
\label{eq12} V = a^{3} = A^{2}B.
\end{equation}
We define the generalized mean Hubble's parameter $H$ as
\begin{equation}
\label{eq13} H = \frac{1}{3}(H_{1} + H_{2} + H_{3}),
\end{equation}
where $H_{1} = \frac{\dot{B}}{B}$  and $H_{2}$ = $H_{3}$ = $\frac{\dot{A}}{A}$ are the directional 
Hubble's parameters in the directions of x, y and z respectively. \\
From Eqs. (\ref{eq12}) and (\ref{eq13}), we obtain
\begin{equation}
\label{eq14} H = \frac{1}{3}\frac{\dot{V}}{V} = \frac{\dot{a}}{a} =
\frac{1}{3}\left(\frac{2\dot{A}}{A} + \frac{\dot{B}}{B}\right).
\end{equation}
An important observational quantity is the deceleration parameter $q$, which is defined as
\begin{equation}
\label{eq15} q = - \frac{\ddot{a} a}{\dot{a}^{2}}.
\end{equation}
The physical quantities of observational interest in cosmology i.e. the expansion scalar $\theta$, 
the shear scalar $\sigma^{2}$ and the average anisotropy parameter $Am$ are defined as
\begin{equation}
\label{eq16} \theta = u^{i}_{;i} = \left(\frac{2\dot{A}}{A} +
\frac{\dot{B}}{B}\right),
\end{equation}
\begin{equation}
\label{eq17} \sigma^{2} =  \frac{1}{2} \sigma_{ij}\sigma^{ij} =
\frac{1}{2}\left[\frac{2\dot{A}^{2}}{A^{2}} + \frac{\dot{B}^{2}}{B^{2}} \right] - 
\frac{\theta^{2}}{6},
\end{equation}
\begin{equation}
\label{eq18} Am = \frac{1}{3}\sum_{i = 1}^{3}{\left(\frac{\triangle
H_{i}}{H}\right)^{2}},
\end{equation}
where $\triangle H_{i} = H_{i} - H (i = 1, 2, 3)$.
\section{Solutions of the Field  Equations}
We have revisited the solution recently obtained by Pradhan et al. (2010). The field equations 
(\ref{eq7})-(\ref{eq9}) are a system of three equations with four unknown parameters $A$, $B$, $\rho$ 
and $\lambda$. One additional constraint relating these parameters are required to obtain explicit 
solutions of the system. We assume that the expansion ($\theta$) in the model is proportional to the
shear ($\sigma$). This condition leads to
\begin{equation}
\label{eq19}
A = B^{m}
\end{equation}
where $m$ is constant. The motive behind assuming this condition is explained with
reference to Thorne \cite{ref56}, the observations of the velocity-red-shift relation for extragalactic 
sources suggest that Hubble expansion of the universe is isotropic today within $\approx 30$ per cent 
\cite{ref57,ref58}. To put more precisely, red-shift studies place the limit
$$
\frac{\sigma}{H} \leq 0.3
$$
on the ratio of shear $\sigma$ to Hubble constant $H$ in the neighbourhood of our Galaxy today. Collins 
et al. \cite{ref59} have pointed out that for spatially homogeneous metric, the normal congruence to the 
homogeneous expansion satisfies that the condition $\frac{\sigma}{\theta}$ is constant. \\\\
Eqs. (\ref{eq8}) and (\ref{eq19}) lead to
\begin{equation}
\label{eq20}
(m + 1)\frac{\ddot{B}}{B} + m^{2} \frac{\dot{B}^{2}}{B^{2}} + \frac{1}{4}B^{2(1-2m)} = 0.
\end{equation}
The general solution of equation (\ref{eq20}) is given by
\begin{equation}
\label{eq21}
t =  \left[ \begin{array}{ll}
            \int{\frac{dB}{k\sqrt{\frac{2r}{s+1}B^{\frac{2k+s-1}{k}}+\alpha B^{\frac{2k-2}{k}}}}} &
 \mbox { when $m\neq\sqrt{2}$}\\
            \int{\frac{dB}{kB^{\frac{k-1}{k}}\sqrt{\frac{2r}{k}ln|B|+\alpha}}}  & \mbox { when $m= \sqrt{2}$} 
            \end{array} \right. 
\end{equation}
where $\alpha$ is a positive integrating constant, $r=-\frac{m^2+m+1}{4(m+1)^2}$, $s=\frac{-3m^2-m+3}
{m^2+m+1}$ and $k=\frac{m+1}{m^2+m+1}$ \\
\subsection{Case(i): when $m\neq \sqrt{2}$}
For $m\neq \sqrt{2}$, equation (\ref{eq20}) leads to
\begin{equation}
\label{eq22}
dt=\frac{dB}{k\sqrt{\frac{2r}{s+1}B^{\frac{2k+s-1}{k}}+\alpha B^{\frac{2k-2}{k}}}}
\end{equation}
Hence the model (\ref{eq1}) is reduced to
\begin{equation}
\label{eq23}
ds^{2} = - \frac{dB^{2}}{k^2\left(\frac{2r}{s+1}B^{\frac{2k+s-1}{k}}+\alpha B^{\frac{2k-2}{k}}\right)}
 + B^{2}(dx + zdy)^{2} + B^{2m}(dy^{2} + dz^{2}).
\end{equation}
After using a suitable transformation of co-ordinates the model (\ref{eq22}) reduces to
\begin{equation}
\label{eq24}
ds^{2} = - \frac{dT^{2}}{k^2\left(\frac{2r}{s+1}T^{\frac{2k+s-1}{k}}+\alpha T^{\frac{2k-2}{k}}\right)} 
+ T^{2}(dx + zdy)^{2} + T^{2m}(dy^{2} + dz^{2})
\end{equation}
where $B=T$ and $dt=\frac{dT}{k\sqrt{\frac{2r}{s+1}T^{\frac{2k+s-1}{k}}+\alpha T^{\frac{2k-2}{k}}}}$.
\subsection{Case(ii): when $m = \sqrt{2}$}
For $m = \sqrt{2}$, equation (\ref{eq20}) lead to
\begin{equation}
\label{eq25}
dt = \frac{dB}{kB^{\frac{k-1}{k}}\sqrt{\frac{2r}{k}\ln{|B|} + \alpha}}
\end{equation}
Hence the model (\ref{eq1}) is reduced to
\begin{equation}
\label{eq26}
ds^{2} = - \frac{dB^{2}}{k^{2}B^{\frac{2(k-1)}{k}}\left(\frac{2r}{k}\ln{|B|} + \alpha\right)} + 
B^{2}(dx + zdy)^{2} + B^{2m}(dy^{2} + dz^{2}).
\end{equation}
After using a suitable transformation of co-ordinates the model (\ref{eq26}) reduces to
\begin{equation}
\label{eq27}
ds^{2} = - \frac{d\tau^{2}}{k^{2}\tau^{\frac{2(k-1)}{k}}\left(\frac{2r}{k}\ln{|\tau|} + \alpha\right)} 
+ \tau^{2}(dx + zdy)^{2} + \tau^{2m}(dy^{2} + dz^{2})
\end{equation}
where $B=\tau$ and $dt=\frac{d\tau}{k\tau^{\frac{k-1}{k}}\sqrt{\frac{2r}{k}\ln{|\tau|} + \alpha}}$.
\section{Some Physical and Geometric Properties of the Models}
Here we discuss some physical and kinematic properties of string model (\ref{eq24}) and model (\ref{eq27}).
\subsection{Case(i): when $m\neq \sqrt{2}$}
The energy density ($\rho$), the string tension $(\lambda)$ and the particle density ($\rho_{p}$) 
for the model (\ref{eq24}) are given by
\[
 \lambda = \frac{3}{4}T^{2(1 - 2m)}- mk^2\left(\frac{2r}{s + 1}T^{\frac{s - 1}{k}} + \alpha 
T^{\frac{-2}{k}}\right)
\]
\begin{equation}
\label{eq28}
- \frac{2mk}{s + 1}\left[(2k + s - 1)kT^{\frac{s-1}{k}} + (k-1)(s+1)\alpha T^{\frac{-2}{k}}\right],
\end{equation}
\begin{equation}
\label{eq29}
\rho = m(m + 2)k^2\left(\frac{2r}{s + 1}T^{\frac{s - 1}{k}} + \alpha T^{\frac{-2}{k}}\right) - 
\frac{1}{4}T^{2(1 - 2m)},
\end{equation}
\[
 \rho_{p} = m(m + 3)k^2\left(\frac{2r}{s + 1}T^{\frac{s - 1}{k}} + \alpha T^{\frac{-2}{k}}\right)
\]
\begin{equation}
\label{eq30}
+\frac{2mk}{s + 1}\left[(2k + s - 1)kT^{\frac{s - 1}{k}} + (k - 1)(s + 1)\alpha T^{\frac{-2}{k}}\right] 
-T^{2(1 - 2m)}.
\end{equation}
From (\ref{eq29}), we observe that $\rho \geq 0$ when
\begin{equation}
\label{eq31}
T^{\frac{4k(m - 1) + 1}{k}}\left(\frac{2r}{s + 1}T^{s + 1} + \alpha\right) \geq \frac{1}{4(m + 2)k^{2}}.
\end{equation}
From (\ref{eq30}), we observe that $\rho_{p} \geq 0$ when
\begin{equation}
\label{eq32}
\frac{2mk^{2}}{(s + 1)}\{r(m + 3) + (2k + s - 1)\} T^{\frac{s - 1}{k}} + m k \alpha (km + 5k - 2)
T^{-\frac{2}{k}} \geq T^{2(1 - 2m)}.
\end{equation}
Hence the energy conditions $\rho \geq 0$ and $\rho \geq 0$ are satisfied under above conditions.\\\\
From Eq. (\ref{eq29}), it is noted that the proper energy density $\rho(t)$ is a decreasing function 
of time and it approaches a small positive value at present epoch. This behaviour is clearly depicted 
in Figure 1 as a representative case with appropriate choice of constants of integration and other 
physical parameters using reasonably well known situations.\\\\
From Eq. (\ref{eq30}), it can be seen that the particle density $\rho_{p}$ is a decreasing function 
of time and $\rho_{p} > 0$ for all times. This nature of $\rho_{p}$ is clearly shown in Figure 3.\\\\ 
All the physical quantities $\rho$, $\rho_{p}$ and $\lambda$ tend to infinity at $T = 0$ and 0 at 
$T = \infty$. The model (\ref{eq25}) therefore starts with a big-bang at $T = o$ and it goes on 
expanding until it comes to rest at $T = \infty$. We also note that $T = 0$ and $T = \infty$ 
respectively correspond to the proper time $t = 0$ and $t = \infty$. There is a point type singularity 
(MacCallum, \cite{ref60}) in the model at $T = 0$. Both $\rho_{p}$ and $\lambda$ tend to zero asymptotically.\\\\ 
We note for $m = \frac{1}{2}$ and $T \to 0$
\begin{equation}
\label{eq33}
\frac{\rho_{p}}{\lambda} = - \frac{4}{3}.
\end{equation}

According to Refs. (see Kibble, \cite{ref1}; Krori et al., \cite{ref61}), when $\rho_{p}/\mid \lambda \mid > 1$, in 
the process of evolution, the universe is dominated by massive strings, and when $\rho_{p}/\mid \lambda
 \mid < 1$, the universe is dominated by the strings. From Eq. (\ref{eq33}), we observe 
\[
\frac{\rho_{p}}{\mid \lambda \mid} < 1 .                                .
\]
Thus, in this case, the universe is dominated by the strings in early stage of evolution of the universe.\\

\begin{figure}
\begin{center}
\includegraphics[width=4.0in]{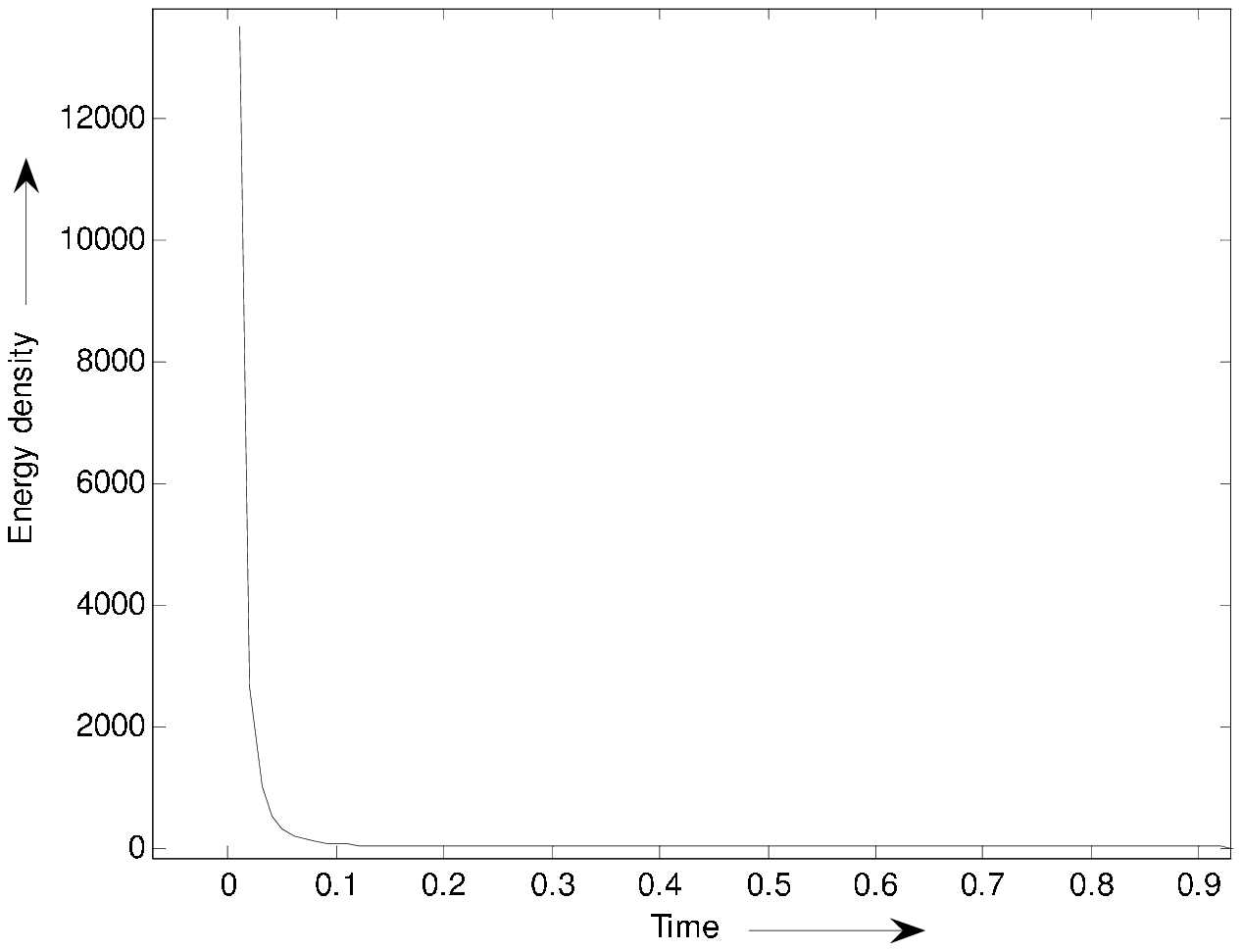} 
\caption{The plot of energy density ($\rho$) vs. time (T)}
\label{fg:archana3fig1.eps}
\end{center}
\end{figure}
\begin{figure}
\begin{center}
\includegraphics[width=4.0in]{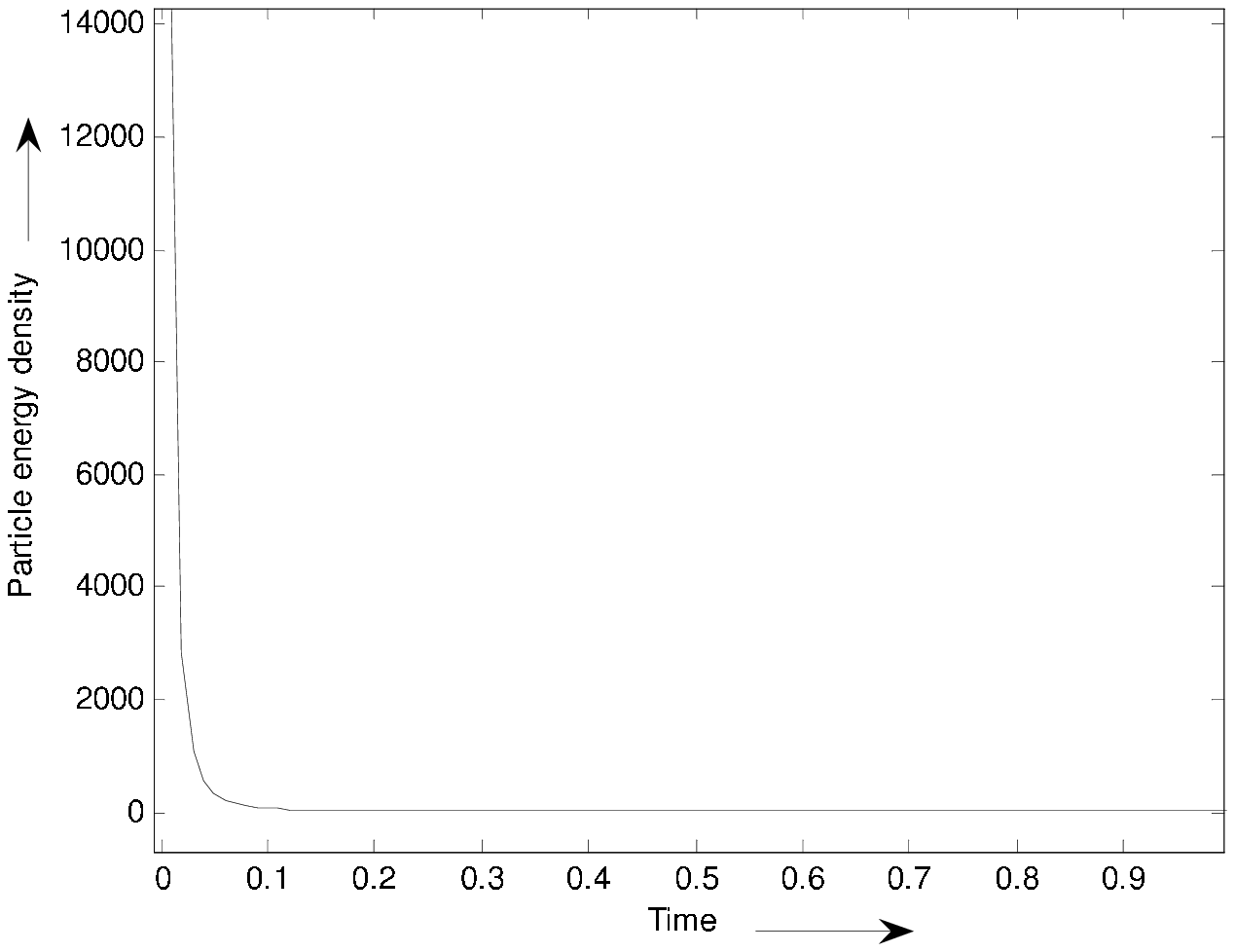} 
\caption{The plot of particle energy density ($\rho_{p}$) vs. time (T)}
\label{fg:archana3fig2.eps}
\end{center}
\end{figure}
The expressions for the scalar of expansion $\theta$, magnitude of shear $\sigma^{2}$, the average 
anisotropy parameter $A_{m}$, deceleration parameter $q$ and proper volume $V$ for the model 
(\ref{eq24}) are given by
\begin{equation}
\label{eq34} \theta = (2m + 1)k\left(\frac{2r}{s+1}T^{\frac{s-1}{k}} + \alpha T^{\frac{-2}{k}}
\right)^\frac{1}{2},
\end{equation}
\begin{equation}
\label{eq35} \sigma^{2} = \frac{(m-1)^2k^2}{3}\left(\frac{2r}{s+1}T^{\frac{s-1}{k}} + 
\alpha T^{\frac{-2}{k}}\right),
\end{equation}
\begin{equation}
\label{eq36} A_{m} =  2\left(\frac{m - 1}{2m + 1}\right)^{2},
\end{equation}
\begin{equation}
\label{eq37} q = - 1 - \frac{3\left[\frac{2r(s-1)}{(s+1)k}T^{\frac{s-1}{k}} - \frac{2\alpha}{k}
T^{\frac{-2}{k}}\right]}{2(2m+1)\left[\frac{2r}{s+1}T^\frac{s-1}{k} - \alpha T^{\frac{-2}{k}}\right]},
\end{equation}
\begin{equation}
\label{eq38} V = T^{2m + 1}. 
\end{equation}
The rate of expansion $H_{i}$ in the direction of x, y and z are
given by
\begin{equation}
\label{eq39} H_{1} = k\left(\frac{2r}{s+1}T^{\frac{s-1}{k}} + \alpha T^{\frac{-2}{k}}\right)^\frac{1}{2}
\end{equation}
\begin{equation}
\label{eq40} H_{2} = H_{3} = mk\left(\frac{2r}{s+1}T^{\frac{s-1}{k}} + 
\alpha T^{\frac{-2}{k}}\right)^\frac{1}{2}
\end{equation}
Hence the average generalized Hubble's parameter is given by
\begin{equation}
\label{eq41} H =  \frac{(2m + 1)k}{3}\left(\frac{2r}{s+1}T^{\frac{s-1}{k}} + \alpha T^{\frac{-2}{k}}
\right)^\frac{1}{2}.
\end{equation}
From equation (\ref{eq34}) and (\ref{eq35}), we obtain
\begin{equation}
\label{42}
\frac{\sigma}{\theta}=\frac{(m-1)}{\sqrt{3}(2m+1)}.
\end{equation}
From Eq. (\ref{eq37}), we conclude the following three cases:
\[
(i) \,\, \mbox {When $T = \left[\frac{s + 1}{r(s - 1)}\right]^{\frac{k}{(s + 1)}}$, \,  $q = - 1$}, 
\]
i.e., we get de Sitter universe. 
\[
(ii) \,\, \mbox{ For $T < \left[\frac{3(s+1)\alpha + (2m+1)(s+1)\alpha}{3(s+1)+2(2m+1)kr}\right]^\frac{k}{s+1}, 
\, q < 0$},
\]
i.e., we obtain accelerating model of the universe. 
\[
(iii) \,\, \mbox{For $ T > \left[\frac{3(s+1)\alpha + (2m+1)(s+1)\alpha}{3(s+1)+2(2m+1)kr}\right]^\frac{k}{s+1}, 
\, q > 0$},
\]
i.e., the model is in decelerating phase. It is remarkable to mention here that though the current observations of 
SNe Ia (Perlmutter {\it et al.} \cite{ref62}$-$\cite{ref64}, Riess {\it et al.} \cite{ref65,ref66}) and CMBR favour 
accelerating models, but both do not altogether rule out the decelerating ones which are also consistent with these 
observations (see, Vishwakarma, \cite{ref67}).\\\\
The geometric properties of the derived model are almost the same of the model as obtained in previous 
paper of Pradhan et al. \cite{ref55}. So we have not discussed these properties here. It is remarkable to 
mention here that for some particular value of m, we can derive the solution of Pradhan et al. \cite{ref55}. 
\subsection{Case(ii): when $m= \sqrt{2}$}
The energy density ($\rho$), the string tension $(\lambda)$ and the particle density ($\rho_{p}$) 
for the model (\ref{eq27}) are given by
\begin{equation}
\label{eq43}
\lambda = \frac{3}{4}\tau^{2(1-2m)} - mk\tau^{\frac{-2}{k}}\left[2r + (2mk-1)\left(\frac{2r}{k}\ln{|\tau|} 
+ \alpha\right)\right],
\end{equation}
\begin{equation}
\label{eq44}
\rho = m(m + 2)k^2\tau^{\frac{-2}{k}}\left(\frac{2r}{k}\ln{|\tau|} + \alpha\right) - \frac{1}{4}\tau^{2(1-2m)},
\end{equation}
\begin{figure}
\begin{center}
\includegraphics[width=4.0in]{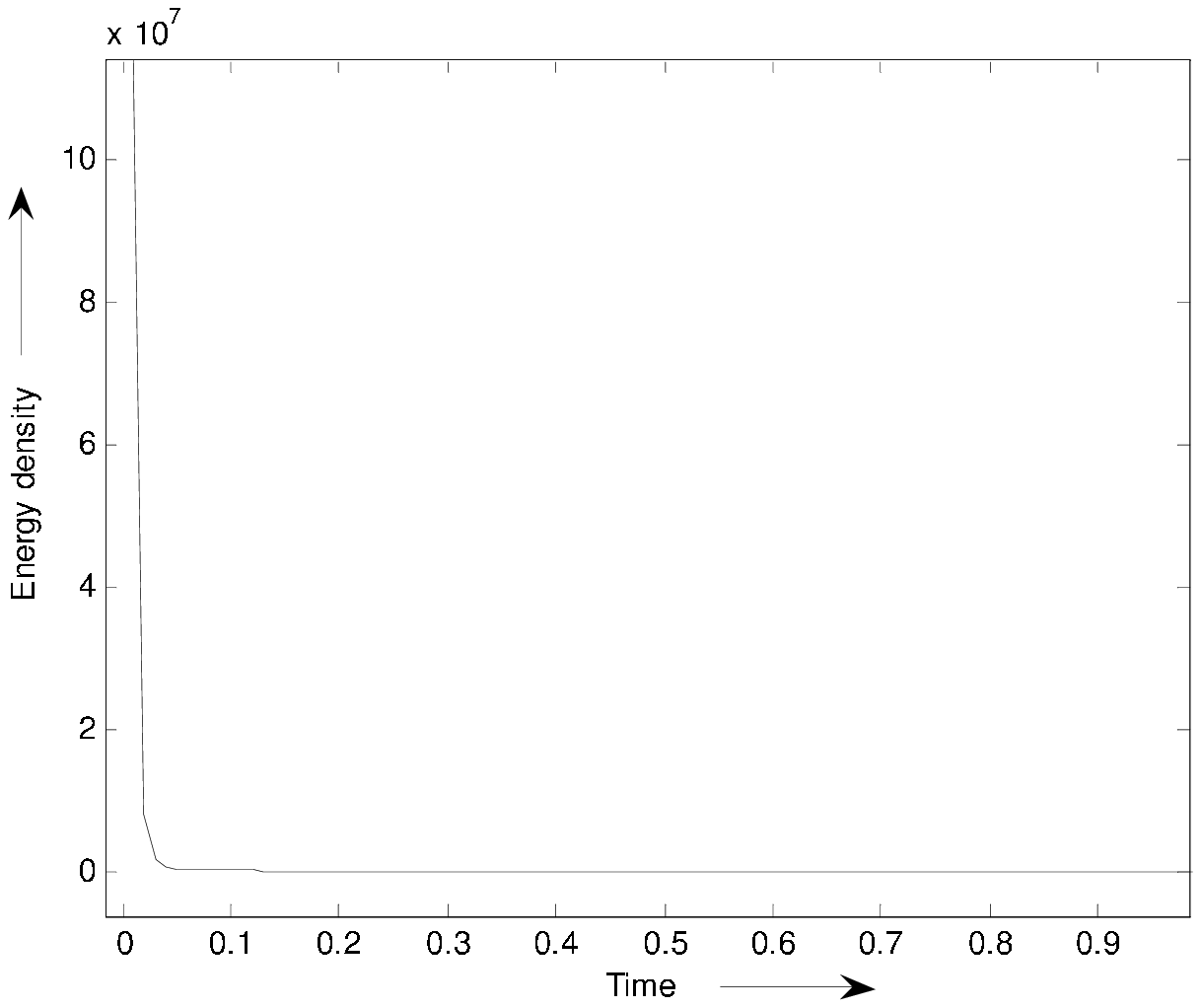} 
\caption{The plot of energy density ($\rho$) vs. time ($\tau$)}
\label{fg:archana3fig3.eps}
\end{center}
\end{figure}
\begin{figure}
\begin{center}
\includegraphics[width=4.0in]{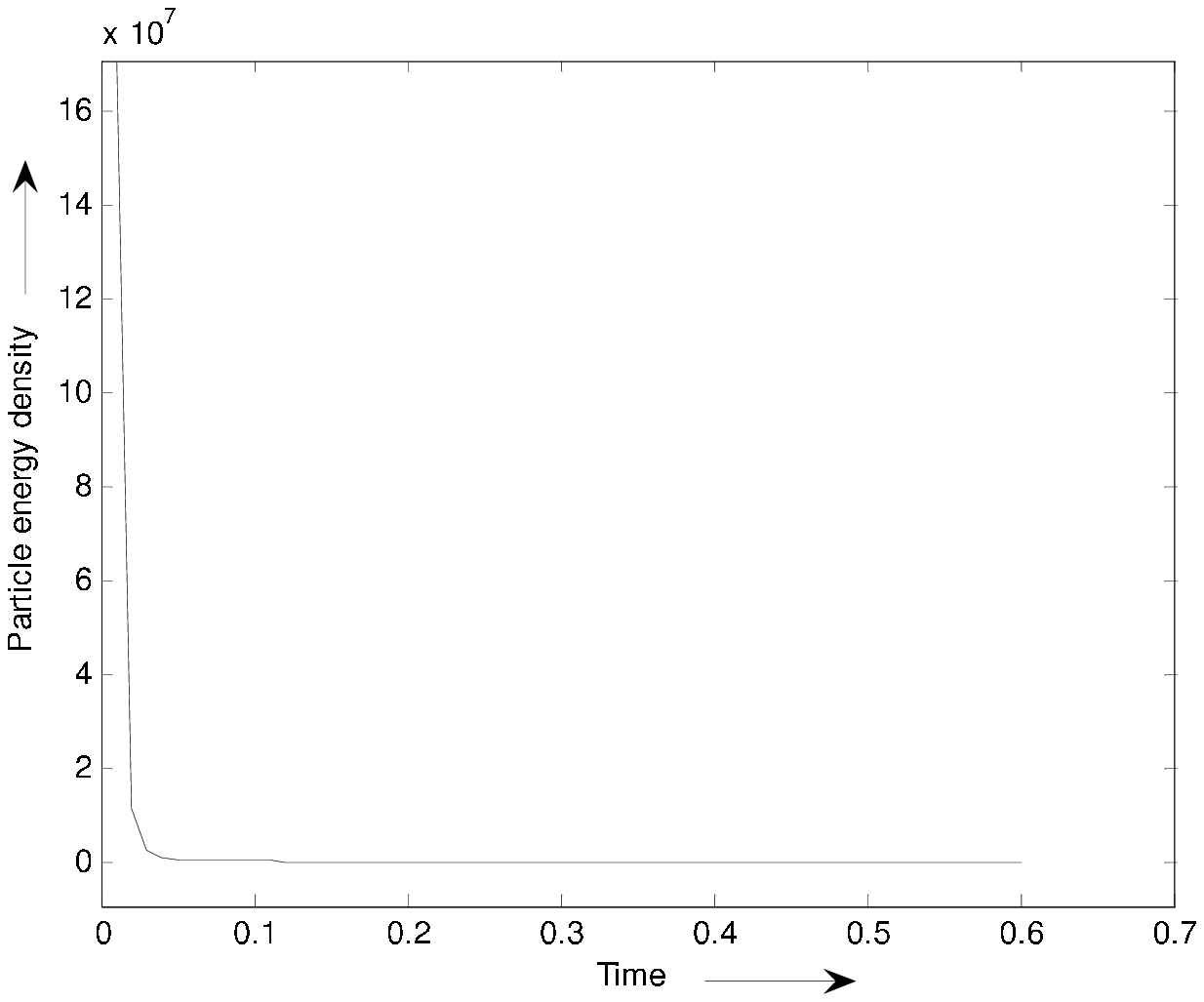} 
\caption{The plot of particle energy density ($\rho_{p}$) vs. time ($\tau$)}
\label{fg:archana3fig4.eps}
\end{center}
\end{figure}

\begin{equation}
\label{eq45}
\rho_{p} = mk\tau^{\frac{-2}{k}}\left[2r + (3mk + 2k - 1)\left(\frac{2r}{k}\ln{|\tau|} + \alpha\right)\right] 
- \tau^{2(1 - 2m)}.
\end{equation}
From (\ref{eq44}), we observe that $\rho \geq 0$ when
\begin{equation}
\label{eq46}
4m(m + 2)k^{2}\left(\frac{2r}{k}\ln{|\tau|} + \alpha \right) \geq \tau^{2(1 - 2m - \frac{1}{k})}.
\end{equation}
From (\ref{eq45}), we observe that $\rho_{p} \geq 0$ when
\begin{equation}
\label{eq47}
mk\left[2r + (3mk + 2k - 1)\left(\frac{2r}{k}\ln{|\tau|} + \alpha \right)\right] \geq \tau^{2(1 - 2m - 
\frac{1}{k})}.
\end{equation}
Hence the energy conditions $\rho \geq 0$ and $\rho \geq 0$ are satisfied under above conditions (\ref{eq46}) 
and (\ref{eq47}).\\\\
From Eq. (\ref{eq44}), it is noted that the proper energy density $\rho(t)$ is a decreasing function 
of time and it approaches a small positive value at present epoch. Figure 3 clearly shows this behaviour of 
$\rho(t)$. \\\\
From Eq. (\ref{eq45}), it is observed that the particle density $\rho_{p}$ is a decreasing function 
of time and $\rho_{p} > 0$ for all times. This nature of $\rho_{p}$ is clearly shown in Figure 4.\\\\ 
All the physical quantities $\rho$, $\rho_{p}$ and $\lambda$ tend to infinity at $\tau = 0$ and 0 at 
$\tau = \infty$. The model is expanding, shearing and non-rotating in general. We also note that $\tau = 0$ 
and $\tau = \infty$ respectively correspond to the proper time $t = 0$ and $t = \infty$. Both $\rho_{p}$ 
and $\lambda$ tend to zero asymptotically.\\\\ 
From Eqs.(\ref{eq45}) and (\ref{eq43}), we note that for  $m = \frac{1}{2}$ and $\tau \to 0$
\begin{equation}
\label{eq48}
\frac{\rho_{p}}{\lambda} = - \frac{4}{3} < 1,
\end{equation}
which shows that the model is dominated by the strings in early stage of evolution of the universe.\\
The expressions for the scalar of expansion $\theta$, magnitude of shear $\sigma^{2}$, the average 
anisotropy parameter $A_{m}$, deceleration parameter $q$ and proper volume $V$ for the model 
(\ref{eq27}) are given by
\begin{equation}
\label{eq49}
\theta = (2m+1)k\tau^{\frac{-1}{k}}\sqrt{\frac{2r}{k}ln|\tau|+\alpha}
\end{equation}
\begin{equation}
\label{eq50}
\sigma^2 = \frac{(m-1)^{2}k^{2}}{3}\tau{\frac{-2}{k}}\left(\frac{2r}{k}ln|\tau|+\alpha\right)
\end{equation}
\begin{equation}
\label{eq51}
A_{m} =  2\left(\frac{m - 1}{2m + 1}\right)^{2},
\end{equation}
\begin{equation}
\label{eq52}
q = - 1 - \frac{3\left[1 - \left(\frac{2r}{k}ln|\tau| + \alpha\right)\right]}{(2m+1)k
\left(\frac{2r}{k}ln|\tau|+\alpha\right)}
\end{equation}
\begin{equation}
\label{eq53}
V =\tau^{2m+1}
\end{equation}
The rate of expansion $H_{i}$ in the direction of x, y and z are
given by
\begin{equation}
\label{eq54}
H_{1} = k\tau{-\frac{1}{k}}\sqrt{\frac{2r}{k}\ln|\tau| + \alpha}
\end{equation}
\begin{equation}
\label{eq55}
H_{2}= H_{3}= mk\tau{-\frac{1}{k}}\sqrt{\frac{2r}{k}\ln|\tau| + \alpha}
\end{equation}
Hence the average generalized Hubble's parameter is given by
\begin{equation}
\label{eq56}
H = \frac{(2m + 1)k}{3}\tau^{\frac{-1}{k}}\sqrt{\frac{2r}{k}\ln|\tau| + \alpha}
\end{equation}
From equation (\ref{eq49}) and (\ref{eq50}), we obtain
\begin{equation}
\label{eq57}
\frac{\sigma}{\theta} = \frac{(m - 1)}{\sqrt{3}(2m + 1)}
\end{equation}
From Eq. (\ref{eq52}), we conclude the following three cases:
\[
(i) \,\, \mbox{When $\ln{|\tau|} = \frac{(1 - \alpha)k}{2r}$, \,  $q = - 1$}, 
\]
i.e. we get de-Sitter universe.\\
\[
(ii) \,\, \mbox{If $\tau < \exp{\left[\frac{k}{2r}\left(\frac{3}{3-(2m+1)k} - \alpha\right) \right]}, \, q < 0$},  
\]
i.e. we obtain accelerating model of the universe.
\[
(iii) \,\, \mbox{For $\tau > \exp{\left[\frac{k}{2r}\left(\frac{3}{3-(2m+1)k} - \alpha\right) \right]} , 
\, q > 0$},
\]
i.e. the model is in decelerating phase. \\\\
From the above results, it can be seen that the spatial volume is zero at $\tau = 0$ and it 
increases with the increase of $\tau$. This shows that the universe starts evolving with zero 
volume at $\tau = 0$ and expands with cosmic time $\tau$. The shear scalar diverges at $\tau = 0$. 
As $ \tau \to \infty$, the scale factors $A(t)$, $B(t)$ tend to infinity. The expansion 
scalar and shear scalar all tend to zero as $\tau \to \infty$. The mean anisotropy parameter is 
uniform throughout whole expansion of the universe when $m \ne - \frac{1}{2}$ but for $m = -\frac{1}{2}$ 
it tends to infinity. The dynamics of the mean anisotropy parameter depends on the value of m. This 
shows that the universe is expanding with the increase of cosmic time but the rate of expansion and 
shear scalar decrease to zero and tend to isotropic.At the initial stage of expansion, when $\rho$ is 
large, the Hubble parameter is also large and with the expansion of the universe $H$, $\theta$ decrease 
as does $\rho$. Since  $\frac{\sigma}{\theta}$ is constant, the model does not approach isotropy at any time.    
\section{Concluding Remarks}
In this paper we have presented two types of new exact solutions of Einstein's field equations for LRS Bianchi type-II 
space-time with a cloud of strings which are different from the other author's solutions. In general, the models are 
expanding, shearing and non-rotating. The energy conditions $\rho \geq 0$, $\rho_{p} \geq 0$ are satisfied under 
proper conditions. It is worth mentioned here that in case (i), $T = 0$ and $T = \infty$ correspond to the proper 
time $t = 0$ and $t = \infty$ respectively. Similarly in case (ii), $\tau = 0$ and $\tau = \infty$ correspond to the 
proper time $t = 0$ and $t = \infty$ respectively. The initial singularity of the model is of the Point Type 
(MacCallum, \cite{ref60}). Our universe starts evolving with zero volume at $ t = 0$ and expands with cosmic time (t). 
We observe that $\frac{\sigma}{\theta}$ is constant in both cases, therefore the models do not approach isotropy 
at any time. The models (\ref{eq24}) and (\ref{eq27}) represent realist models. It is interesting to mention here that 
for $m = 1$ , the anisotropy parameter $A_{m}$ tends to zero, therefore the space-time anisotropy disappear and the 
models of our universe approach isotropy for $m = 1$. \\\\
The particle density and the tension density of the string are comparable at the two ends and they fall off 
asymptotically at similar rate. It is observed that in early stage of the evolution of the universe, 
the universes in both cases are dominated by strings. Our solution of case (i) generalizes the solution 
recently obtained by Pradhan et al. \cite{ref55}. The results of this paper are in favour of observational 
features of the universe. \\\\
Finally, we would like to mention here that in the derived models, the proper energy density $\rho$ and particle 
density $\rho_{p}$ are positive. Therefore, the weak energy condition (WEC) as well as null energy condition (NEC) 
are obeyed in the present system.  

\section*{Acknowledgments}
Authors (A. K. Yadav and A. Pradhan) would like to thank the Institute of Mathematical Sciences (IMSc.), 
Chennai, India for providing facility and support under associateship scheme where part of this work was 
carried out. 

\end{document}